\documentclass[aps,prd,longbibliography,
,groupedaddress, twocolumn,notitlepage,superscriptaddress]{revtex4-1}

\usepackage{amsmath,amssymb}
\usepackage{natbib}
\usepackage{graphicx}
\usepackage{hyperref}
\usepackage{subfigure}
\newcommand{\be}{\begin{equation}}
\newcommand{\ee}{\end{equation}}
\newcommand{\bse}{\begin{subequations}}
\newcommand{\ese}{\end{subequations}}
\newcommand{\ba}{\begin{eqnarray}}
\newcommand{\ea}{\end{eqnarray}}
\newcommand{\bea}{\begin{eqnarray}}
\newcommand{\eea}{\end{eqnarray}}

\usepackage{color}
\usepackage[normalem]{ulem}  

\begin{document}


\title{A simple model for strange metallic behavior}


\author{Sutapa Samanta}
\email[]{samants2@wwu.edu}
\affiliation{Department of Physics and Astronomy, Western Washington University,
516 High Street, Bellingham, Washington 98225,USA}
\author{Hareram Swain}
\email[]{dhareram1993@physics.iitm.ac.in}
\affiliation{Department of Physics, Indian Institute of Technology Madras, Chennai 600036, India}
\author{Beno\^{i}t Dou\c{c}ot}
\email[]{doucot@lpthe.jussieu.fr}
\affiliation{Laboratoire  de  Physique  Th\'{e}orique  et  Hautes  Energies,Sorbonne Universit\'{e} and CNRS UMR 7589, 4 place Jussieu, 75252 Paris Cedex 05, France}
\author{Giuseppe Policastro}
\email[]{giuseppe.policastro@ens.fr}
\affiliation{Laboratoire de Physique de l’École Normale Supérieure, ENS, Université PSL, CNRS, Sorbonne Université, Université Paris Cité, 24 rue Lhomond, F-75005 Paris, France}

\author{Ayan Mukhopadhyay}
\email[]{ayan@physics.iitm.ac.in}
\affiliation{Department of Physics, Indian Institute of Technology Madras, Chennai 600036, India}



\date{\today}

\begin{abstract}

A refined semi-holographic non-Fermi liquid model, in which carrier electrons hybridize with operators of a holographic critical sector, has been proposed recently for strange metallic behavior. The model, consistently with effective theory approach, has two couplings whose ratio is related to the doping. We explain the origin of the linear-in-T resistivity and strange metallic behavior as a consequence of the emergence of a universal form of the spectral function which is independent of the model parameters when the ratio of the two couplings take optimal values determined only by the critical exponent. This universal form fits well with photoemission data of copper oxide samples for under/optimal/over-doping with a fixed exponent over a wide range of temperatures. We further obtain a refined Planckian dissipation scenario in which the scattering time $\tau = f \cdot \hbar /(k_B T)$, with $f$ being $\mathcal{O}(1)$ at strong coupling, but $\mathcal{O}(10)$ at weak coupling.

\end{abstract}

\maketitle
\section{Introduction}
\noindent An elegant explanation for universal unconventional transport phenomena of strange metals is based on the existence of a quantum critical point (QCP) at optimal doping \cite{Sachdev:2011cs,Taillefer,Michon2019,Hayes2021}. It has been further emphasized via many tractable models that the quantum critical degrees of freedom may also provide the key to the emergence of unconventional superconducting and insulating phases. The spectral function measured via angle-resolved photoemission spectroscopy (ARPES) indeed provides evidence for absence of quasi-particles at the Fermi surface \cite{RevModPhys.75.473,Vishik_2010,reber2015power,PhysRevB.78.035103,Lee_2018,RevModPhys.92.031001}. Of late, there has been remarkable progress in achieving direct access to the quantum critical point via the application of a strong magnetic field \cite{2012QCP} and also a critical current \cite{2018criticalcurrent}. However, it has been argued that an effective theory for universal transport properties of strange metals should also incorporate the degrees of freedom at intermediate scales, and which interacts with the infrared critical sector. This hypothesized feature is called \textit{Mottness} \cite{PHILLIPS20061634,PhysRevB.77.014512,PhysRevLett.102.056404,PhysRevLett.106.016404,PhysRevB.83.214522,PhysRevB.86.115118}. 

Another influential paradigm for understanding transport phenomena in strange metals is \textit{Planckian dissipation} \cite{Zaanen2004,sachdev_2011,Bruin2013,Hartnoll:2016apf,Chowdhury:2021qpy,Hartnoll:2021ydi}. Here, the absence of quasi-particles is assumed to coincide with the Mott-Ioffe-Regel (MIR) limit in which the scattering length approaches the interatomic spacing. It is also assumed that all intrinsic scales disappear as the system is close to a QCP. Therefore, a \textit{universal} strong interaction limit emerges where the thermal energy $k_B T$ is the only energy scale, {and therefore the  dissipation time-scale should be $\mathcal{O}(\hbar/(k_B T))$}. This readily accounts for the linear-in-$T$ dc resistivity at low temperatures. However, as discussed later, the abundant experimental data for heavy fermion compounds indicate \cite{Taupin2022AreHF} that this naive version of the paradigm may fail.

In this work, we analyze a simple effective model of strange metallic behavior proposed in \cite{Doucot:2020fvy} that captures features of Mottness. It incorporates lattice band electrons hybridizing with fermionic operators of a quantum critical sector, and can be viewed as an effective theory of the conducting electrons of the lower Hubbard band with only two couplings. It was shown that there always exists an \textit{optimal ratio} of these two couplings at which the model exhibits universal transport properties, especially linear-in-T resistivity, over a very wide range of temperatures, \textit{irrespective} of all parameters provided the critical exponent lies within a certain range. {By \textit{optimal ratio}, we simply imply that the ratio of couplings needs to be tuned such that we obtain linear-in-T resistivity. Here we will show that the fundamental origin of this feature in this model is the emergence of a universal form of the spectral function as a function of the frequency near the Fermi surface over a large range of temperatures. As discussed below, this universal behavior emerges at finite (non-vanishing) temperatures due to a competition between two different types of interactions, and this cannot be obtained from the self-energy contributions due to interactions with the critical sector alone. We also emphasize that the universal form is needed even at large frequencies to reproduce strange-metallic transport properties (although it is valid only near the Fermi surface) due to the underlying non-Fermi liquid nature of the system.}


The critical sector of this model can be best viewed as a homogeneous bath of Sachdev-Ye-Kitaev (SYK) quantum dots \cite{Sachdev:1992fk,Kitaev:2017awl}, each of which has a scaling symmetry in the large $N$ and infrared limit, and admits a dual holographic description in the form of a black hole in two-dimensional anti-de Sitter ($AdS_2$) space. The fermionic operators of this emergent holographic infrared conformal field theory (IR-CFT) are dual to free Dirac electrons living in the $AdS_2$ black hole spacetimes whose masses determined by the dual scaling dimensions \cite{Lee:2008xf,Liu:2009dm,Cubrovic:2009ye,Faulkner:2009wj,Faulkner:2010da,Iqbal:2011in}. 

The model is a special case of the more general semi-holographic semi-local non-Fermi liquids constructed in \cite{Mukhopadhyay:2013dqa} as extension of the simple Polchinski-Faulkner setup \cite{Faulkner:2010tq} in which only linear hybridization of the lattice electrons with fermionic operators of the critical sector was considered. In \cite{Mukhopadhyay:2013dqa,Doucot:2017bdm} it was shown that, in a suitable large $N$ limit, the low frequency behavior of the spectral function on the Fermi surface is given only by the linear hybridization considered in the Polchinski-Faulkner scenario even after incorporating arbitrary interactions between lattice electrons, and also between lattice electrons and the holographic IR CFT sector; thus implying the existence of \textit{generalized quasi-particles} on the Fermi surface.

In \cite{Doucot:2020fvy}, it was proposed that as a special case of \cite{Mukhopadhyay:2013dqa,Doucot:2017bdm}, one can consider two irrelevant couplings on the Fermi surface which give rise to two distinct \textit{local} self-energy terms, one of which is Fermi-liquid like and another given by the finite temperature fermionic holographic Green's function. Since the latter term originates from the bath of SYK-type quantum dots, {and we assume a smeared homogeneous distribution of these dots, the corresponding coupling is related to the density of these dots, and thus the doping strength.} 

Here we will explain the origin of the strange metallic behavior seen in \cite{Doucot:2020fvy} from the emergence of a universal form of the spectral function {as a function of the frequency} near the Fermi surface, which is independent of the temperature, the critical exponent and other model parameters, and which also fits nodal ARPES data well. Furthermore, we will obtain a refined Planckian dissipation paradigm from our effective approach in consistency with experimental data.

We will use units where $\hbar = k_B =1$.

\section{The refined semi-holographic model}
\noindent In the model proposed in \cite{Doucot:2020fvy}, the conducting electron ($\hat{c}$) interacts with a fermionic operator $\hat\chi_{\rm CFT}$ of the holographic IR-CFT, and also with other filled lattice band electrons ($\hat{f}$). Crucially, all interactions scale with $N$ in a specific way. The Hamiltonian has the following generic form in the large $N$ limit: 
\begin{align}\label{modelhamiltonian}
\hat{H} = &\sum_\mathbf{k} \epsilon(\mathbf{k}) \hat{c}^\dagger (\mathbf{k})\hat{c} (\mathbf{k}) + N\sum_\mathbf{k} \left( g \hat{c}^\dagger(\mathbf{k}) \hat\chi_{\text{CFT}} (\mathbf{k}) + c.c. +\cdots\right)\nonumber\\
& + N^2 \hat{H}_{\text{IR CFT}} + \sum_{i,j,k} \left(\lambda_{ijk, \mathbf{k}_1, \mathbf{k}_2, \mathbf{k}_3 } \right.\\
&\left. \hat{c}^\dagger(\mathbf{k}_1) \hat{f}_i (\mathbf{k}_2) \hat{f}_j^\dagger(\mathbf{k}_3) \hat{f}_k(\mathbf{k}_1 - \mathbf{k}_2 + \mathbf{k}_3)+ c.c\right) + \cdots ,\nonumber
\end{align}
in which interactions involving $c$ that are multi-linear in $\chi_{\rm CFT}$ are suppressed, those which are linear in $\chi_{\rm CFT}$ are $\mathcal{O}(N)$, and those which do not involve $\chi_{\rm CFT}$ are $\mathcal{O}(N^0)$. We have removed $c^\dagger f$ type interactions by diagonalizing the $\mathcal{O}(N^0)$ quadratic terms. The above is a special case of the semi-holographic models analyzed in \cite{Mukhopadhyay:2013dqa,Doucot:2017bdm} implying that we can infer its low energy behavior via Wilsonian RG approach in the large $N$ limit. 

The propagating degree of freedom on the Fermi surface is essentially $c$ plus a $\mathcal{O}(1/N)$ contribution from $\chi$ \cite{Mukhopadhyay:2013dqa}. The resulting retarded Green's function on the Fermi surface takes the following form in the large $N$ limit (see the supplemental material of~\cite{Doucot:2020fvy} for a derivation {by considering perturbative corrections in $\lambda$ and other couplings}):
\begin{align}\label{greensfunction}
G_R(\omega,\mathbf{k}) = &\Bigg( \omega + i \tilde{\gamma} (\omega^2 +\pi^2 T^2) +\tilde{\alpha} \mathcal{G}(\omega)\nonumber\\
& - \tfrac1{2m}(\mathbf{k}^2 - k_F^2)+\mathcal{O}\left(\frac{1}{N}\right)\Bigg)^{-1}.
\end{align}
Here $\tilde{\gamma} \sim \mathcal{O}(\lambda^2)$ and $\tilde{\alpha} \sim \mathcal{O}(g^2)$ are the two leading (irrelevant) couplings determining the Fermi-liquid and holographic components of the self-energy respectively with the latter given by~\cite{Iqbal:2009fd}
\begin{align}\label{adspropagator}
 \mathcal{G}(\omega) = e^{i(\phi + \pi\nu/2)} (2\pi T)^\nu \frac{\Gamma\left(\tfrac12 +\tfrac{\nu}2 -i\tfrac{\omega}{2\pi T}\right)}{\Gamma\left(\tfrac12 -\tfrac{\nu}2 -i\tfrac{\omega}{2\pi T}\right)} \, .
\end{align}
In order that the spectral function $\rho = - 2 \text{Im } G_R$ is non-negative, we require $0< \phi < \pi(1-\nu)$. We choose $\phi\approx \pi(1-\nu)/2$ as discussed in \cite{Doucot:2020fvy} so that (only) at the Fermi momentum, the spectral function has particle-hole symmetry. However, we do not need significant fine-tuning of $\phi$ for our conclusions to hold.

It is convenient to re-write the propagator~\eqref{greensfunction} with variables re-scaled with respect to the Fermi energy scale ($E_F =k_F^2/2m$) as follows:
\begin{align}\label{rescaledpropagator}
G_R(x,\mathbf{y}) =& E_F^{-1} \left( x+ i \gamma(x^2 + (\pi x_T)^2) +\alpha e^{i(\phi +\pi\nu/2)}(2\pi x_T)^\nu \right.\nonumber\\
&\left. \frac{\Gamma\left(\tfrac12 +\tfrac{\nu}2 -i\tfrac{x}{2\pi x_T}\right)}{\Gamma\left(\tfrac12 -\tfrac{\nu}2 -i\tfrac{x}{2\pi x_T}\right)} - (\vert\mathbf{y}\vert^2 -1) \right)^{-1}\, .
\end{align}
Above $x=\omega/E_F$, $x_T = T/E_F$, $\mathbf{y} = \mathbf{k}/k_F $, $\gamma= \tilde{\gamma}E_F$, and $\alpha = \tilde{\alpha}E_F^{-(1-\nu)}$. Therefore, the dimensionless effective couplings of the model are $\alpha$ and $\gamma$. For this effective description to be valid at temperatures smaller than the Fermi energy, we require both of these couplings to be small.

The ratio $\alpha/\gamma$ is naturally related to the doping strength following our previous discussion. The Fermi energy is the only scale of the model, however the scale at which this model should be matched to the microscopic description must be lower.

\section{Universality and fits to ARPES data}

\noindent In \cite{Doucot:2020fvy} it was shown that for any value of $\nu$ in the range $0.66 \lessapprox \nu\lessapprox0.95$, there exists an optimal ratio $\alpha/\gamma$ at which the model \ref{modelhamiltonian} exhibits linear-in-T d.c. resistivity over a very wide range of temperatures. At this optimal ratio (for its dependence on $\nu$ see plots in \cite{Doucot:2020fvy}), the Fermi liquid and holographic self-energy contributions finite temperature self-energy contributions shown in \ref{greensfunction} (and \ref{rescaledpropagator}) balance out to produce new scaling behavior. Here we show that the key to such strange metallic behavior is the emergence of a universal form of the spectral function.

Let us define new variables $\tilde{x} = \omega/T =x/x_T$ and $\tilde{\mathbf{y}}$ via $ \vert\tilde{\mathbf{y}}\vert^2 -1 = x_T^{-1} \left(\vert\mathbf{y}\vert^2 - 1\right)$. Note $\vert\tilde{\mathbf{y}}\vert = 1$ corresponds to the Fermi surface. Furthermore, we define the function $F$ from the retarded propagator \eqref{rescaledpropagator} as follows:
\begin{align}\label{Eq:Spec-Universal}
&F(\tilde{x},\tilde{\mathbf{y}}):= T G_R(\omega,\mathbf{k}) = T G_R(\tilde{x},\tilde{\mathbf{y}}) 
\nonumber\\
&=\Big( \tilde{x}+ i \gamma x_T(\tilde{x}^2 + \pi^2) +\alpha x_T^{-1}(2\pi x_T)^\nu \nonumber\\
&\hspace{0.5cm} e^{i(\phi +\pi\nu/2)}
 \frac{\Gamma\left(\tfrac12 +\tfrac{\nu}2 -i\tfrac{\tilde{x}}{2\pi}\right)}{\Gamma\left(\tfrac12 -\tfrac{\nu}2 -i\tfrac{\tilde{x}}{2\pi }\right)} - (\vert\tilde{\mathbf{y}}\vert^2 -1) \Big)^{-1}.
\end{align}
{By universality, we mean that $-2 T {\rm Im} G_R(\omega,\mathbf{k}) = T\rho(\omega,\mathbf{k}) = -2 {\rm Im}F(\tilde{x},\tilde{\mathbf{y}})$ is approximately the \textit{same} function of $\tilde{x}$ (i.e. $\omega/T$) when (i) $\vert \tilde{\mathbf{y}}\vert \sim 1$ and fixed (i.e. the momentum is near the Fermi surface) and (ii) the ratio of the couplings $\alpha/\gamma$ is in a narrow \textit{optimal} range of values corresponding to the choice of the scaling exponent $\nu$. The explicit dependence on $x_T = T/E_F$ thus disappears in the universal regime to a good approximation, and so does the explicit dependence on the scaling exponent $\nu$, and also the overall scale of the couplings (as only their ratio matters). We find that the universality emerges only at finite temperatures for $t_c < x_T (=T/E_F) < 10$, and also when $0.66 \lessapprox \nu\lessapprox0.95$ and when both couplings are small ($\gamma\lessapprox 0.01$). The lower end of $x_T$, i.e. $t_c$ depends on the choice of $\nu$. For $\nu \approx 0.66$, $t_c \sim 0.01$. However, for higher values of $\nu \approx 0.95$, universality emerges at much lower temperatures $x_T (=T/E_F)\sim 0.01$. We emphasize that both the optimal ratio $\alpha/\gamma$ and $t_c$ depend only on $\nu$, and not on the overall scale of the couplings (i.e. $\gamma$) in quantitative agreement with \cite{Doucot:2020fvy}.}

{The origin of universality is the competition of the two finite temperature effects in the local self-energy contributions, one which is Fermi-liquid like and another which is holographic. It is also important to note that although unievrsality emerges only near the Fermi surface, it should be valid for the full range of frequencies, as the tail of the spectral function also contributes to transport properties, as we will see soon. It is important that we are at finite temperature, as otherwise $\omega^2$ (the Fermi-liquid contribution) and $\omega^{2\nu}$ (the holographic contribution) in the self-energy, cannot compete with each other at arbitrary values of the frequency.}

\begin{figure}[ht]
\includegraphics[width=0.45\textwidth]{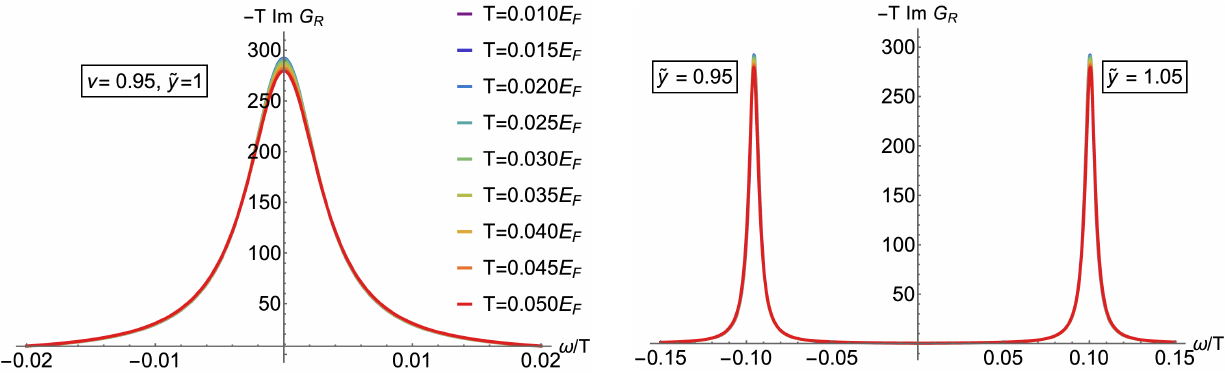}
\caption{Top: Plot of $(1/2)T \rho$ as a function of $\omega/T$ for various $T \geq 0.01 E_F$ at the optimal $\alpha/\gamma = 100$ for $\nu = 0.95$ on the Fermi surface (left), as well as near it (right). In all plots, $\gamma = 0.001$.}\label{Fig:spectraluniversality}
\end{figure}
\begin{figure}[h]
\includegraphics[width= 0.27\textwidth]{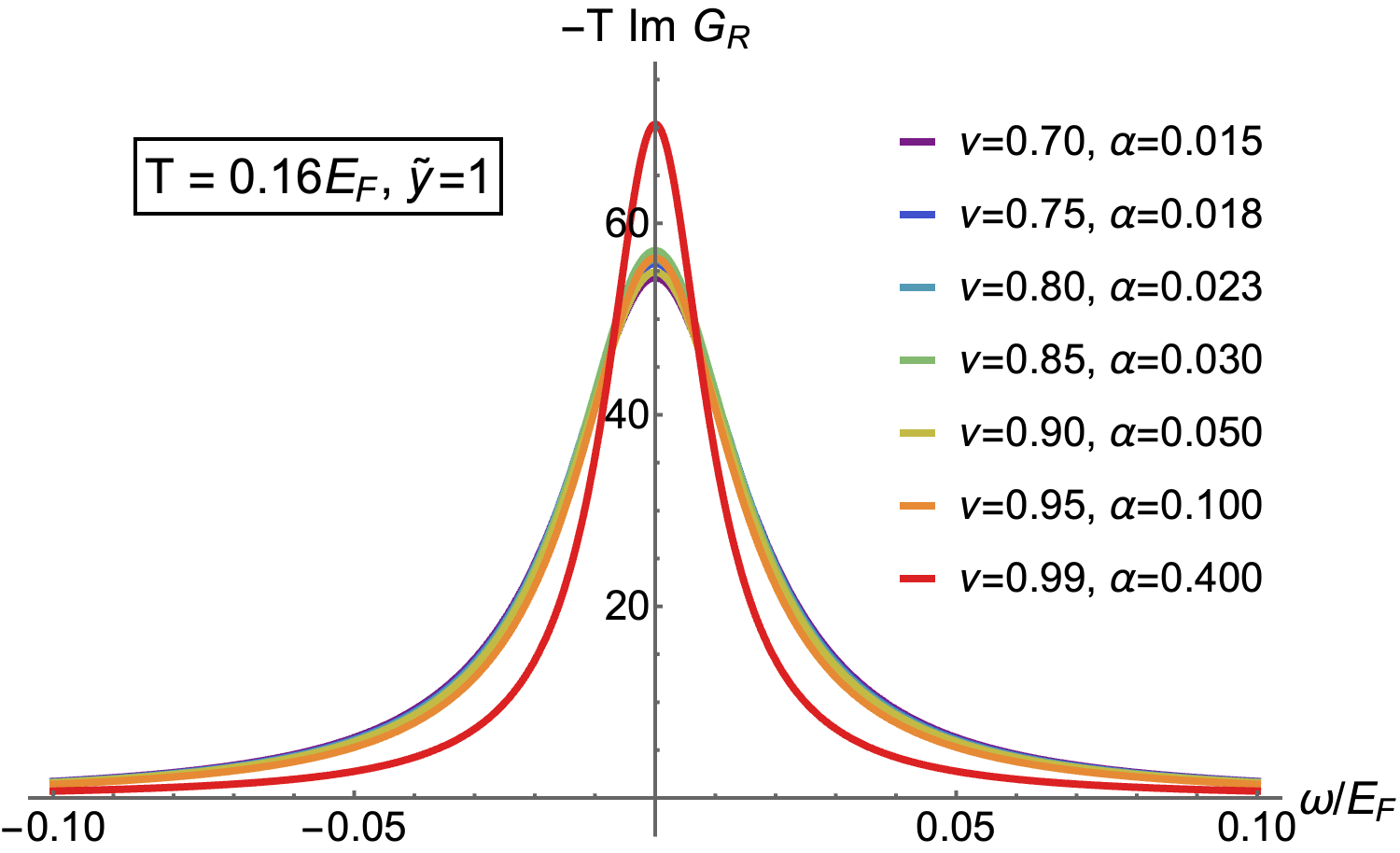}
\includegraphics[width= 0.27\textwidth]{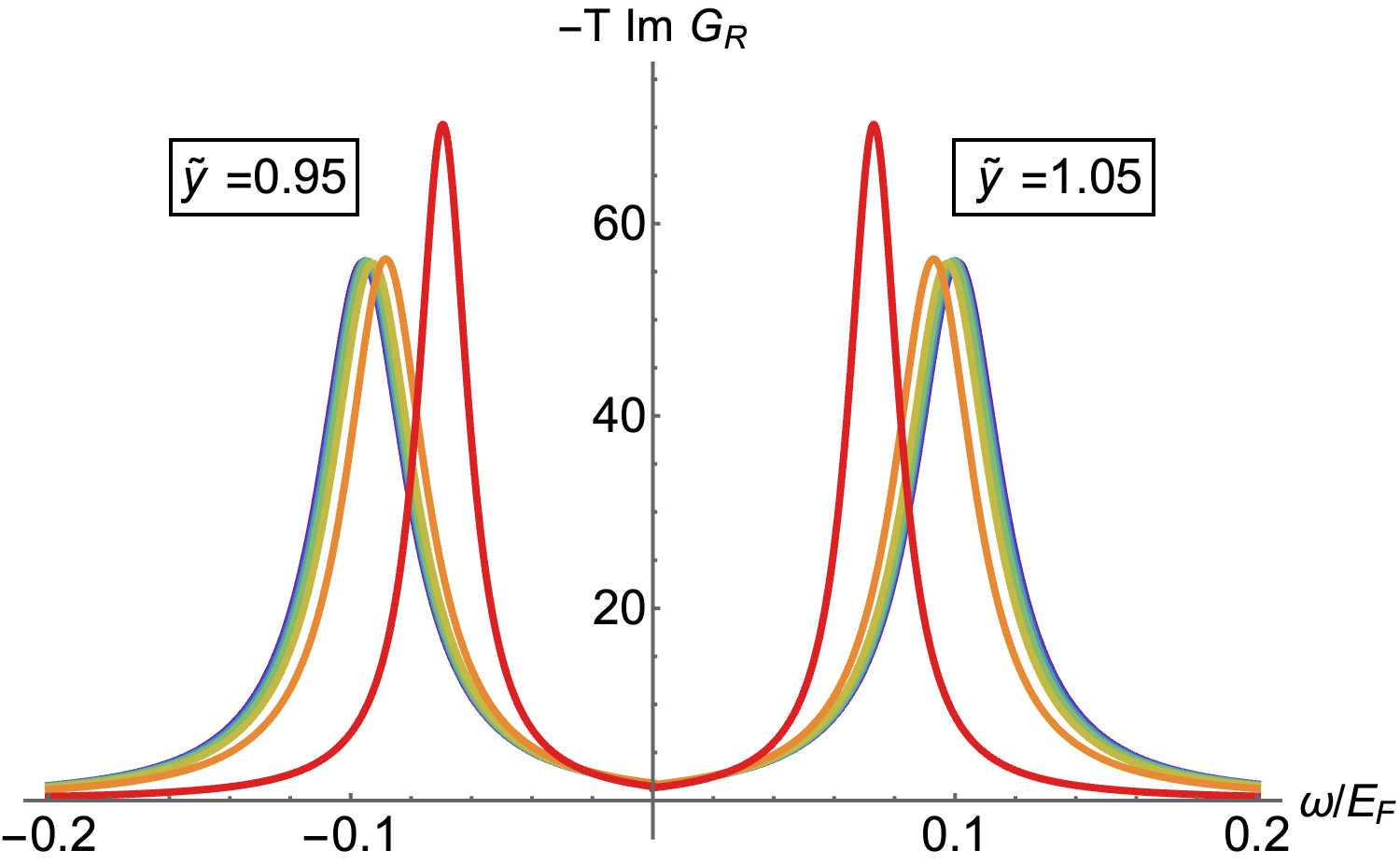}
\caption{Universality for different values of $\nu$ on the Fermi surface (top) as well as near it (bottom) for corresponding optimal $\alpha/\gamma$, and $\gamma = 0.001$. We have chosen $T=0.16E_F$.}\label{spectraluniversalityWnu}
\end{figure}

We can readily see from Fig~\ref{Fig:spectraluniversality} that as a function of the frequency, the scaled spectral function $T(\rho/2) = -  T {\rm Im} G_R = -  {\rm Im} F$ is independent of the temperature to an excellent approximation, for $T > 0.01 E_F$ when plotted as a function of $\tilde{x} = \omega/T$ on the Fermi surface ($\vert\tilde{\mathbf{y}}\vert = 1$) and also near it ($\vert\tilde{\mathbf{y}}\vert = 0.95$ and $\vert\tilde{\mathbf{y}}\vert = 1.05$), at the optimal ratio $\alpha/\gamma =100$ corresponding to $\nu =0.95$, as claimed above. Furthermore, from plots in Fig. \ref{spectraluniversalityWnu}, we find that $T(\rho/2)$ is also independent of $\nu$ at the corresponding optimal ratios $\alpha/\gamma$ (at a representative temperature $T=0.16 E_F$) both at the Fermi surface and near it.  Together these plots validate that $F$ is indeed independent of $x_T$ and also the model parameters like $\nu$ in the universal regime.

\begin{table}[h]
\centering
\begin{tabular}{|c|c|c||c|c|c|}
\hline
 \multicolumn{3}{|c||}{  Optimal }  &\multicolumn{3}{c|} {Non-Optimal}   \\
\hline
\hline
    $\nu$ & $\alpha$ & $\delta$ & $\nu$ & $\alpha$ & $\delta$  \\ \hline
    0.95 & 0.100 & 7.0 & 0.95 & 0.200 & 30.0 \\ \hline 
    0.90 & 0.050 & 2.7 & 0.90 & 0.100 & 36.8 \\ \hline
    0.85 & 0.030 & 3.3 & 0.85 & 0.100 & 60.7 \\ \hline
    0.75 & 0.018 & 0.8 & 0.80 & 0.050 & 43.5 \\ \hline
    0.70 & 0.015 & 2.8 & 0.99 & 0.400 & 34.6 \\ 
\hline
\end{tabular}
\caption{Left: $ \delta$ for ``optimal set of values" of $\alpha$ for given choices of $\nu$ and $\gamma =0.001$. We see that $\delta <10$. Right: The same for some set of values which are ``not optimal" for which $\delta$ is significantly larger than $10$. }
\label{universalitytablefig2}
\end{table}

{Although the universality is well demonstrated in Figs. ~\ref{Fig:spectraluniversality} and ~\ref{spectraluniversalityWnu} by how well the spectral functions multiplied by the temperature collapse on to a single curve, a more quantitative study is also illuminating. To develop a quantitative measure, we first define a standard curve, 
\begin{widetext}
\begin{equation}
    f_s := (-2 T {\rm Im} G_R(\omega, \mathbf{k}))[\nu = 0.8, \alpha = 0.023, \gamma = 0.001, T/E_F = 0.16],
\end{equation}
\end{widetext}
which is in the universal regime, Furthermore, let 
\begin{equation}
    \frac{\delta}{100} = \frac{\int_{-\infty}^\infty \vert(-2 Im F - f_s)\vert {\rm d}\tilde x}{\int_{-\infty}^\infty F_s {\rm d}\tilde x},
\end{equation}
where $F$ is defined in \eqref{Eq:Spec-Universal}. A simple quantitative test of universality of $F$ is then to check if $\delta < 10$.}


\begin{table}[h]
\centering
\begin{tabular}{|c|c|c||c|c|c|}
\hline
 \multicolumn{3}{|c||}{  Optimal }  &\multicolumn{3}{c|} {Non-Optimal}   \\
\hline
\hline
    $\nu$ & $\gamma$ & $\delta$ & $\nu$ & $\gamma$ & $\delta$  \\ \hline
    0.95 & 0.009 & 3.0 & 0.95 & 0.002 & 57.9 \\ \hline 
    0.90 & 0.006 & 3.8 & 0.90 & 0.001 & 40.9 \\ \hline
    0.85 & 0.004 & 2.1 & 0.85 & 0.009 & 26.7 \\ \hline
    0.87 & 0.005 & 2.1 & 0.87 & 0.001 & 26.4 \\ \hline
\end{tabular}
\caption{$\delta$ for  optimal and non-optimal values of $\gamma$ for corresponding values of $\nu$ with $\alpha =0.023$ and $T/E_F = 0.16$ fixed. We see that $\delta <10$ for optimal cases, while for non-optimal cases $\delta$ is much larger than $10$. This also provides evidence that the universality is determined only by the ratio of $\alpha/\gamma$.}
\label{non-optimal gamma}
\end{table}
{In Table ~\ref{universalitytablefig2}, we have studied the values of $\delta$ with variation of $\alpha$ for various fixed values of $\nu$ at $x_T = T/E_F =0.16$ both in and away from the optimal regime. In Table ~\ref{non-optimal gamma}, we have studied the values of $\delta$ when $\gamma$ is varied from the optimal value at various fixed values of $\nu$ while keeping $\alpha =0.023$ and $x_T =T/E_F = 0.16$ fixed as in the standard curve.  We readily note that in the optimal regime, $\delta$ is typically much smaller than $10$ while in the non-optimal regime it is significantly larger. As we keep $\alpha$ instead of $\gamma$ fixed in Table ~\ref{non-optimal gamma} unlike in our plots and Table ~\ref{universalitytablefig2}, the two tables together provide evidence that universality is determined only by the ratio $\alpha/\gamma$ (at the corresponding $\nu$) and not by their absolute values (although we need $\gamma<0.01$ for the universality to emerge).}

{Particularly, Table ~\ref{non-optimal gamma} is also remarkable because we see that, although $\gamma$ is small, a small variation in $\gamma$ can produce a significant departure from universality. Therefore, the effect of $\gamma$ is significant not only at infinitesimal temperatures (recall that in this table $x_T = T/E_F=0.16$). At this point, it is appropriate to emphasize again that $\delta <10$ provides a good physical test of universality because the contribution to transport comes also from higher frequencies, which constitute the tail of the spectral function, as will be discussed below.}

{Furthermore, one can similarly verify quantitatively by computing $\delta$ explicitly that the universality emerges only when $t_c < x_T (=T/E_F) < 10$, and that $t_c$ is determined by $\nu$ alone.}


\subsection{Comparing with ARPES data}
\begin{figure}[h]
\includegraphics[width=0.47\textwidth]{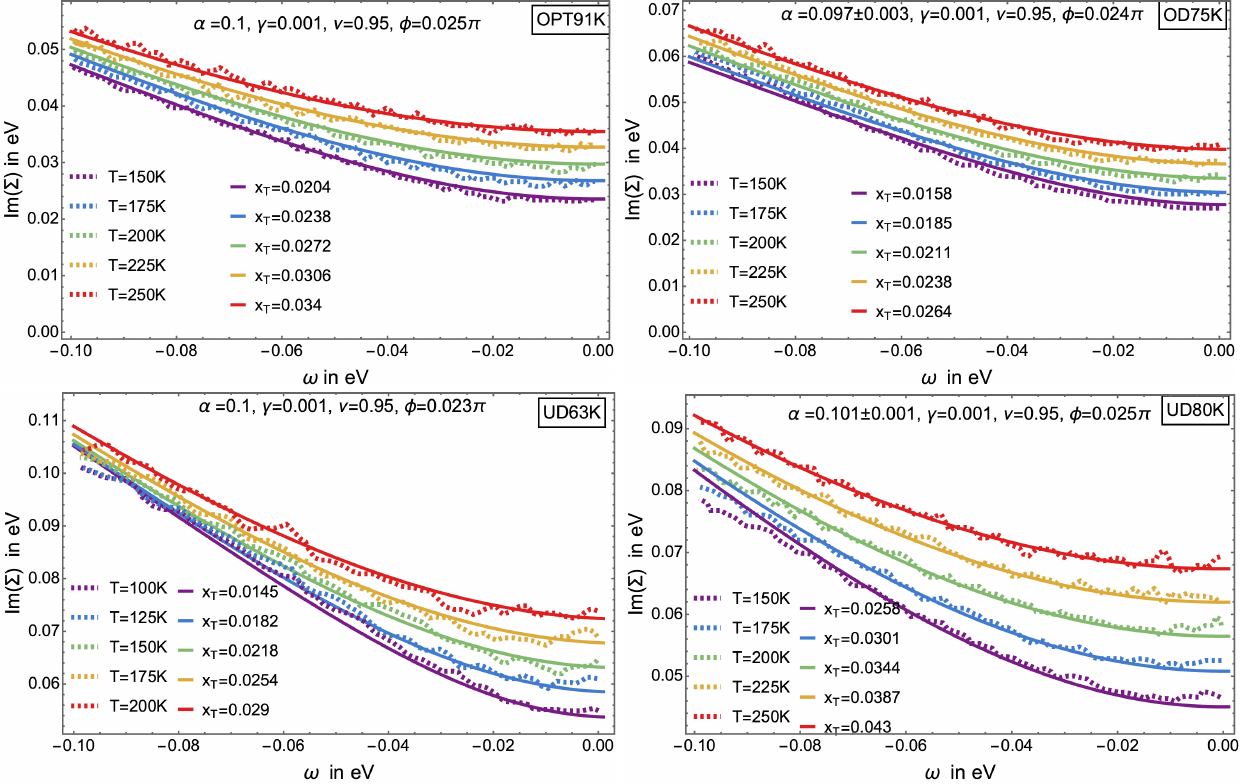}
\caption{Fit of the imaginary part of self energy with the ARPES data from \cite{Reber2019}. The labels OPT, OD and UD denote optimal/over/under-doping and the following numbers denote $T_c$ in K.}\label{selfenergy}
\end{figure}

\noindent As shown in Fig.~\ref{selfenergy}, our spectral function fits well with the \textit{nodal} ARPES data for the imaginary part of the self-energy for different underdoped, overdoped and optimally doped samples of $Bi_2Sr_2CaCu_2O_{8+\delta}$ at various temperatures, reported in \cite{Reber2019}. The fits work well keeping the Fermi energy and the value of $\alpha$ fixed for each doping (for all temperatures), and with $\gamma =0.001$ chosen in all cases \footnote{The data has been extracted form the paper using WebPlotDigitizer tool. \url{https://automeris.io/WebPlotDigitizer/}.}. The exponent $\nu = 0.95$ is remarkably stable across doping, and the fitted value of $\alpha$ indeed corresponds to the optimal ratio at which we obtain universality. 
However, $\alpha$ does decrease very mildly in the overdoped region implying more Fermi-liquid like behavior. In \cite{Reber2019}, the fits were obtained with varying exponents and different phenomenological ansatz which was improved in \cite{Smit} from theoretical considerations. However, our fits work well with more constraints.

{We have not done a quantitative study of goodness of fit due to the significant noisiness of the data, but our fitting is physically motivated. For the universality to emerge at the temperatures reported in the plot (assuming $E_F \sim 10^4K$ ), we need $\nu$ to be closer to $0.95$ as discussed above. The choice $\nu=0.95$ is then at the edge of universality. Furthermore, we expect $\alpha$ to depend on the doping strength, as discussed before, and not $\gamma$. Therefore, the fits with varying $\alpha$ and fixed $\gamma$ suggest the validity of the underlying physical intuition for the model.}

{Although our model needs to include order parameters for description of various phases, the very mild variation of $\alpha$ and hence the ratio of couplings with the doping in our fits could suggest the that strange metallic behavior can be restored to a good approximation even away from optimal doping if order parameters are suppressed. Indeed restoration of strange metallic behavior by suppression of charged density waves has been reported for $YBa_2 Cu_3 O_{7-\delta}$ in underdoped samples below the pseudogap temperature \cite{Restored}. See \cite{Overdoped,Husain} for the overdoped regime.}


\section{Transport and Planckian dissipation}
\noindent {Ignoring vertex corrections and assuming that charge transport is primarily due to the propagating degrees of freedom on the Fermi surface, the dc conductivity can be readily obtained from the spectral function via ($n_F$ is the Fermi-Dirac distribution):}
\begin{align}\label{dcconductivity}
\hspace{-0.27cm}\sigma_{\text{dc}} = \frac{e^2}{m^2}\int \frac{d\omega}{2\pi} \int \frac{d^2 k}{4\pi^2} k^2 \rho(\omega, \mathbf{k}, T)^2 \left(-\dfrac{\partial n_F(\omega, T)}{\partial \omega}\right).
\end{align}

The linear-in-T behavior of the dc resistivity over a wide range of temperatures at the optimal ratio of the couplings simply follows from the universal form of the spectral function \eqref{Eq:Spec-Universal} given that the above integral gets its major contribution from momenta near the Fermi surface. The latter implies that  $k^2$ in the integrand can be replaced by $k_F^2$. To see this, we can replace $\rho$ by $(-1/2T){\rm Im} F(\tilde{x},\tilde{\mathbf{y}})$ following \eqref{Eq:Spec-Universal} in the above integral. A factor of $T^2$ comes from changing the loop frequency and momenta variables to $\tilde{x} = \omega/T$ and $\tilde{\mathbf{y}}\approx \hat{n}k_F(1 + (E_F/T)((k/k_F) -1))$ respectively (with $\hat{n} = \mathbf{k}/k$). An additional factor of $T^{-1}$ arises from the $\omega$-derivative of $n_F$. The integral thus behaves as $1/T$ 
for the very wide range of temperatures over which the universal form of the spectral function \eqref{Eq:Spec-Universal} is valid (e.g. for $T>0.01 E_F$, $\alpha/\gamma \approx 100$ and $\nu \approx 0.95$). 

\begin{figure}[h]
\centering
\begin{subfigure}
  \centering
  \includegraphics[width= 0.232\textwidth]{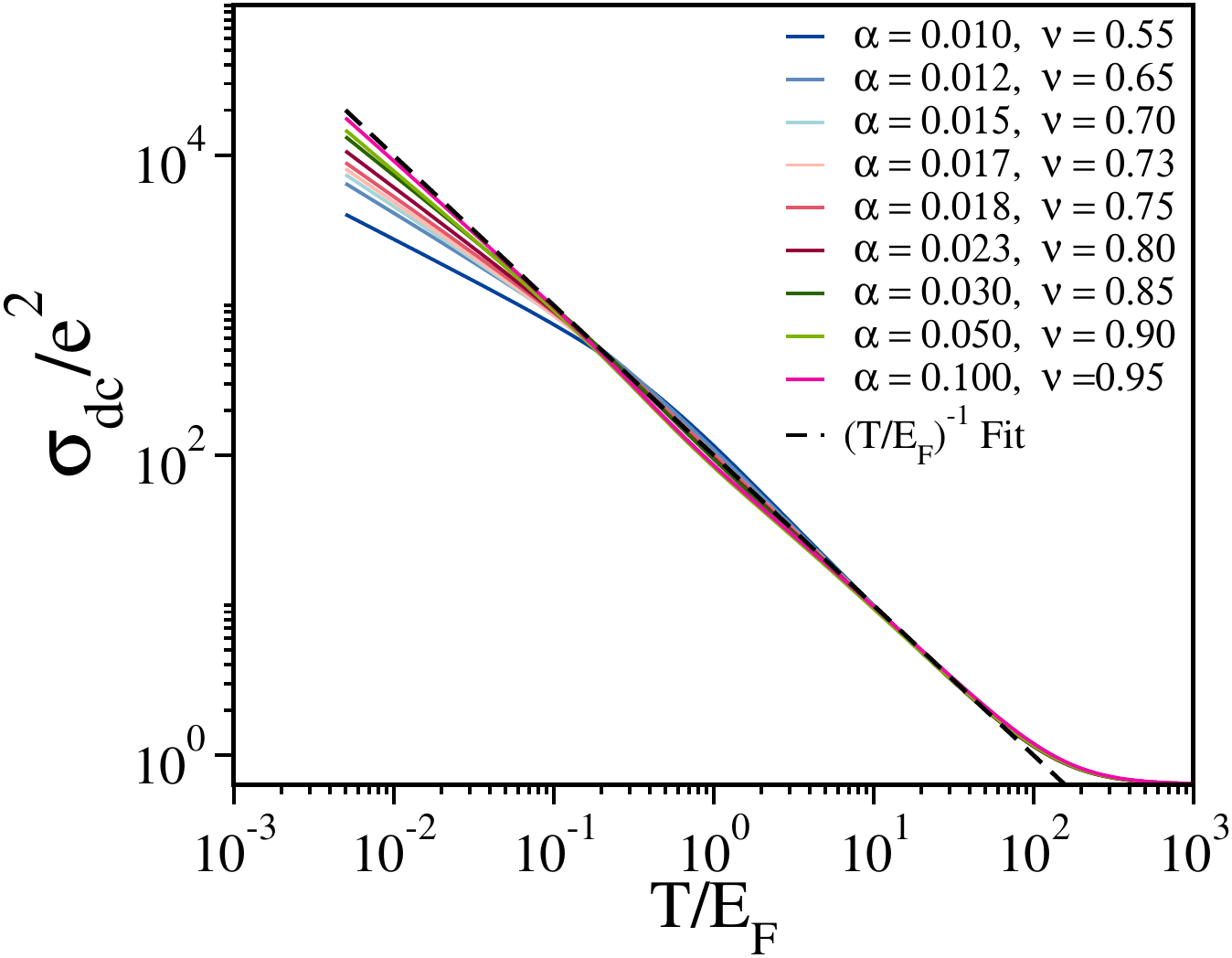}
  \label{fig:sub1}
\end{subfigure}%
\begin{subfigure}
  \centering
  \includegraphics[width=.232\textwidth]{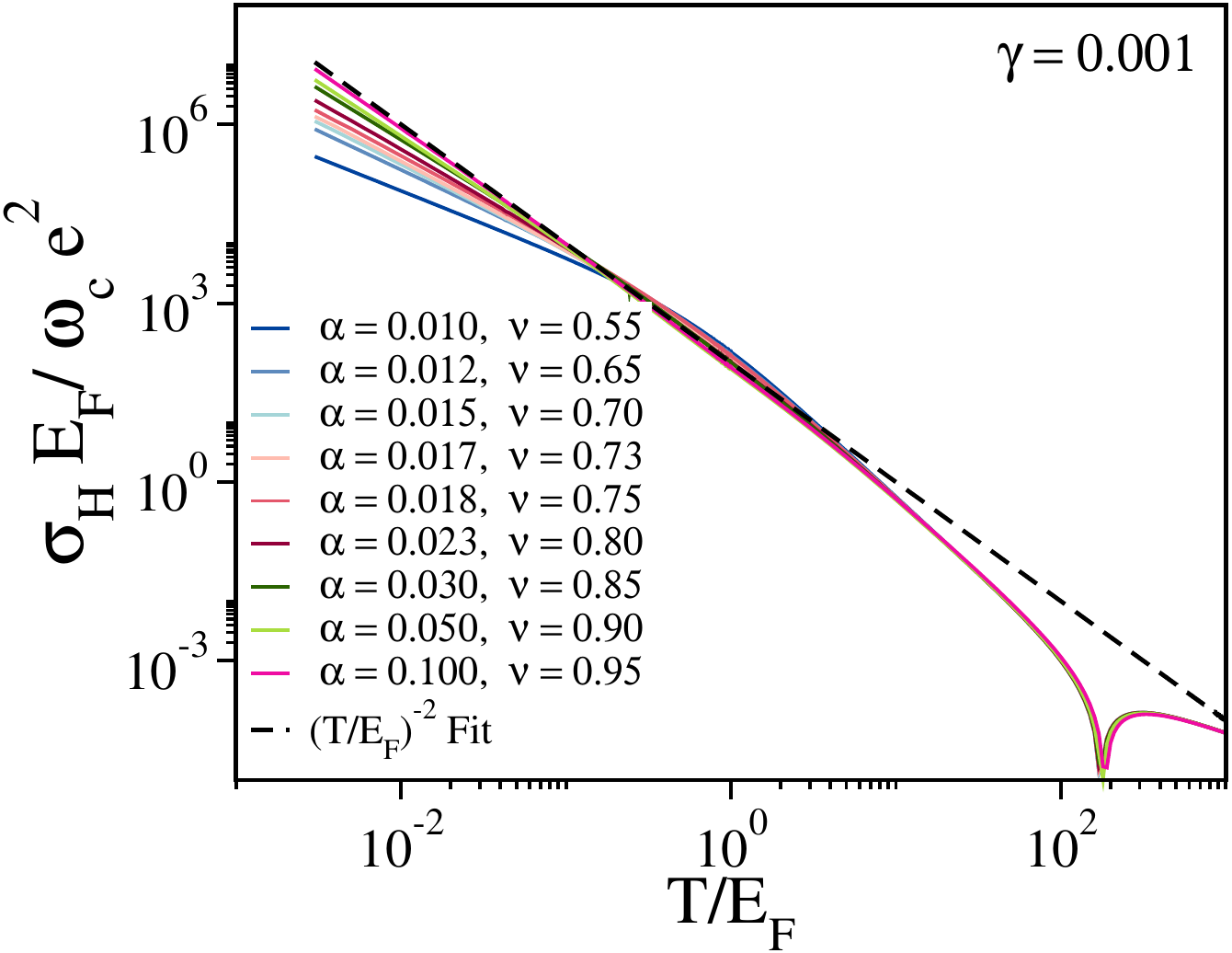}
  \label{fig:sub2}
\end{subfigure}
\caption{The dc conductivity (top) and Hall conductivity (bottom) show $T^{-1}$ and $T^{-2}$ behaviors respectively at the optimal ratios of $\alpha/\gamma$. Here $\gamma =0.001.$}\label{conductivity}
\end{figure}

However, explicit computations shown in Fig.~\ref{conductivity} imply that the linear-in-$T$ behavior of the resistivity holds only up to an upper value of temperature because the integral over loop momenta gets contributions eventually from momenta far away from the Fermi surface where universality breaks down. In reality, the integral over loop momenta has cutoff due to the bandwidth, 
and therefore the linear-in-$T$ scaling should hold to a higher approximation. We note that since the linear-in-T resistivity is obtained only when the integral gets major contribution from the Fermi surface, \textit{our assumption, that only the propagating degrees of freedom on the Fermi surface contribute to the charge transport, is consistent.}

{We also emphasize that although the integral \eqref{dcconductivity} receives contribution from momenta only near the Fermi surface, the large frequency tail of the spectral function contributes significantly. Therefore, for the analytic argument for the linear-in-T resistivity to be valid, the universal form of the spectral function should be a good approximation at all frequencies, as stated before.}

The Hall conductivity $\sigma_H$ for small magnetic fields similarly takes the form ($\omega_c = eB/m$):
\begin{align}\label{hallconductivity}
&\sigma_{\rm H} = {2\omega_c \frac{e^2}{m}} \int \frac{d\omega}{2\pi} \int \frac{d^2 k}{4\pi^2} k_x \rho(\omega,\mathbf{k},T) \dfrac{\partial n_F(\omega, T)}{\partial \omega}\nonumber\\
&\hspace{3cm}\times\left( \dfrac{\partial}{\partial k_x} \text{Re } G_R (\omega,\mathbf{k}, T) \right) \ .
\end{align}
The above readily follows from considering the real time version of the corresponding expression obtained in~\cite{Fukuyama1969,Itoh_1985} (see Appendix \ref{Sec:AppA} for reproduction of the Fermi liquid result). Using the Kramers-Kronig relation for ${\rm Re}\,G_R$, one can argue similarly to the case of the dc-resistivity, that $T^{-2}$ behavior of Hall conductivity should hold if the momentum integral receives major contribution from near the Fermi surface, and \eqref{Eq:Spec-Universal} is valid.

Here, we have assumed a spherical Fermi surface. However, when the Fermi surface has equal amounts of positive and negative curvatures, the integral \eqref{hallconductivity} can be vanishingly small, and the Hall conductivity could behave as $T^{-3}$ instead of $T^{-2}$ as observed in \cite{PhysRevB.68.094502}. We leave this for a future investigation.

Fig.~\ref{conductivity} shows the log-log plots of dc conductivity and Hall conductivity as a function of temperature at the optimal ratios of the couplings corresponding to various $\nu$, and with $\gamma = 0.001$. At low temperatures, the dc and Hall conductivities behave as $T^{-\nu}$ and $T^{-2\nu}$ respectively \footnote{However higher loops contribute significantly at very low $T$ giving residual resistivity, etc.}. Crucially, there is a wide range of temperatures with the lower end coinciding with that of the universal regime (thus starting from $0.01 E_F$) for higher $\nu$) for which the d.c. and Hall conductivities behave as $T^{-1}$ and $T^{-2}$, respectively. For higher $\gamma$, the temperature range over which the universal scaling behavior is obtained, decreases although the lower end of this range does not change (recall the discussion in the previous section that the latter is determined only by $\nu$ and not the scale of couplings).  

 To make contact with Planckian dissipation, we recall the Fermi liquid results obtained from \eqref{dcconductivity} and \eqref{hallconductivity} with Fermi liquid type spectral functions (see \cite{Fukuyama1969} and Appendix \ref{Sec:AppA} for derivations):
 \begin{equation}\label{Eq:Drude1}
\sigma_{\rm dc} = \frac{n e^2}{m} b\,\tau,\quad \sigma_{\rm H} = \frac{ne^2}{m}\omega_c\tau^2,
\end{equation}
where $\tau$ is the scattering time, $b$ is the renormalization factor for the density of states at the Fermi surface and $n$ is the carrier density. 
For the Fermi liquid, $n/m = E_F/\pi$. Motivated by the observation that the integrals \eqref{dcconductivity} and \eqref{hallconductivity} get their major contributions from the Fermi surface in the regime where they show $T^{-1}$ and $T^{-2}$ dependence, we can verify whether \eqref{Eq:Drude} can be consistent with our results although they are not derivable from a quasi-particle picture. Setting $b=1$, \eqref{Eq:Drude1} yields
\begin{equation}\label{Eq:Drude}
\tau = \frac{\sigma_{\rm H}}{\omega_c\sigma_{\rm dc}}, \,\,\,\, \frac{n}{m} = \frac{\sigma_{\rm dc}^2\omega_c}{e^2\sigma_{\rm H}}.
\end{equation}
These can be used to extract $\tau$ and the effective $n/m$ from the explicit $\sigma_{\rm dc}$ and $\sigma_{\rm H}$ obtained from \eqref{dcconductivity} and \eqref{hallconductivity}, respectively, with the spectral function obtained from \eqref{greensfunction}.

\begin{figure}[h]
 \includegraphics[width=0.48\textwidth]{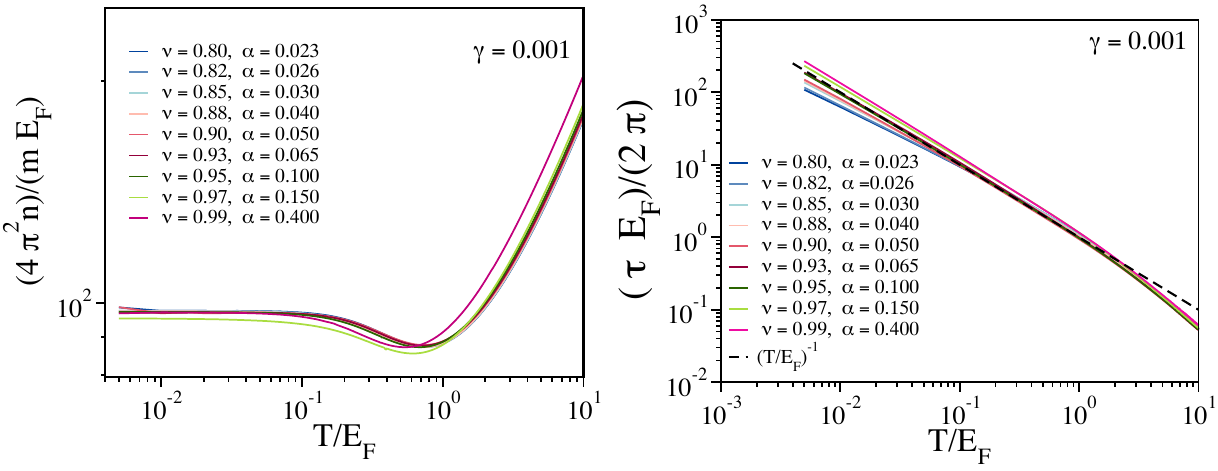}
\caption{The plots of effective $4\pi^2 n/(m E_F)$ (left) and $\tau E_F/(2\pi)$ (right) as functions of $T/E_F$ are shown for various values of $\nu$ (at $\gamma =0.001$) and the corresponding optimal ratios $\alpha/\gamma$. 
}\label{CDandST}
\end{figure}

Plots for $\tau$ and effective $n/m$ as functions of $T/E_F$ are shown in  Fig.~\ref{CDandST} for $\gamma = 0.001$, and for the optimal ratios $\alpha/\gamma$ corresponding to various values of $\nu$. Indeed, $\tau \approx f\cdot T^{-1}$ (in units $\hbar = k_B =1$) with $f$ in the range $2\pi\times 0.9- 2\pi\times 1.3$, and is thus approximately independent of $\nu$ and other details of the critical sector. Planckian behavior with such higher value of $f \approx \mathcal{O}(10)$ can indeed be found in many heavy fermion compounds according to \cite{Taupin2022AreHF}. Furthermore, the effective $n/m$ is approximately constant for a very wide range of temperatures and also lies in the range $2\pi -2.5\pi$ times the corresponding Fermi liquid value ($= E_F/\pi$) as we vary $\nu$ from $0.66$ to $0.95$, confirming a Drude-like phenomenology. 

The value of $f$ decreases as we increase $\gamma$ keeping the optimal ratio $\alpha/\gamma$ fixed, while the effective $n/m$ remains almost invariant. When $\gamma = 0.01$, $f\approx 1$ as found in cuprates \cite{Taupin2022AreHF}. However, $f$ cannot be made smaller than $\mathcal{O}(1)$, as increasing $\gamma$ beyond $0.01$ drastically shrinks the range of temperatures over which universality of the spectral function holds and strange metallic behavior is exhibited. So we naturally obtain the expected strong coupling Planckian limit $f\approx1$. At smaller $\gamma$ and therefore for weaker coupling, $f$ increases as pointed above, in consistency with the analysis of \cite{Taupin2022AreHF}. In contrast, we find that the effective $n/m$ is almost independent of the overall strength of the couplings at their optimal ratio.
\section{Discussion}
\noindent For our simple model to be more realistic and to capture the various phases of cuprates and other materials, we should consider a lattice of $AdS_2$ black holes (dual to the quantum dots) which couple to the itinerant electrons via a critical fermionic operator, include Yukawa type interactions with bulk scalars, and analyze the system with a novel dynamical mean field theory (DMFT) type formulation. This approach is similar to those in \cite{PhysRevLett.119.216601,PhysRevX.8.021049,PhysRevX.8.031024,PhysRevLett.121.187001,Cha_2020,PhysRevLett.122.186601,PhysRevLett.123.066601,PhysRevResearch.2.033434,Patel:2022gdh} and also to quantum black hole microstate models discussed in \cite{Kibe:2020gkx,Kibe:2021gtw}. In the latter case, the strange metallic regime can emerge naturally when the throats behave as decoupled units leading to a local self-energy for the itinerant electrons. The decoupling of the throats has been shown to be necessary for the emergence of black hole complementarity without encountering information paradoxes \cite{Kibe:2020gkx,Kibe:2021gtw,Kibe:2023mzg} in the microstate models \footnote{{In such microstate models, after perturbations due to infalling matter, the throats decouple from each other absorbing energy from infalling matter, while the latter also decouples from the throats \cite{Kibe:2023mzg}. Furthermore, the decoupled residue of the infalling matter (absorbed in the hair) is a time-dependent quantum state which non-isometrically encodes its initial state. The ringdown of the decoupled throats also encodes the same information transitorily and transfers the same to Hawking radiation. Thus information is replicated (but not cloned) as black hole complimentarity demands. The decoupling of all degrees of freedom provides the underlying mechanism for the black hole complimentarity to emerge and in our present context could be relevant for the robustness of the locality of the self-energy.}}

Our simple model with only two effective couplings, nevertheless, provides an effective theory to understand many features of strange metallic transport and a refined version of Planckian dissipation. A deeper understanding could arise from applications of quantum information theory to the lattice models.

 \begin{acknowledgments}
We thank Subhasis Ghosh for comments on the manuscript, and also for very stimulating and helpful discussions. AM acknowledges support from the Ramanujan Fellowship of the Science and Engineering Board (SERB) of the Department of Science and Technology of India. HS acknowledges support from the INSPIRE PhD Fellowship of Department of Science and Technology of India. AM and GP also acknowledge support from IFCPAR/CEFIPRA funded project no 6304-3. SS acknowledges support by the US National Science Foundation under Grant DMR-1945395.
\end{acknowledgments}
\appendix
\section{The conductivities of the Fermi liquid}\label{Sec:AppA}

\noindent In order to derive the Fermi liquid dc and Hall conductivities, we start from the Green's function (as parametrized in \cite{Fukuyama1969})
\begin{align}\label{Eq:FLG}
G_{R} (\omega,\mathbf{k}) = \left(a^{-1} \omega - b^{-1}\epsilon_{k} + i \tau^{-1} \right)^{-1},
\end{align}
where $\tau$ is the scattering time, $a$ and $b$ are renormalization factors. The specific heat particularly is renormalized by $b/a$ 
and $b$ is simply the factor renormalizing the density of states at the Fermi surface. To see the latter, we note that when $\tau$ is large, we can approximate
\begin{align}\label{Eq:FLrho1}
&\rho(0,\mathbf{k}) = - 2 {\rm Im}\,G_{R} (0,\mathbf{k}) \approx 2\pi\delta(b^{-1}\epsilon_{k} ) \nonumber\\
&= 2\pi b \delta(\epsilon_k).
\end{align}

In order to do the integral Eq.~\eqref{dcconductivity}, we proceed at follows. Firstly, at zero temperature
\begin{align}\label{zerotempfermidirac}
\frac{\partial}{\partial \omega} n_{F}(\omega) = - \delta({\omega} ).
\end{align}
Secondly, since the integral gets contribution Eq.~\eqref{dcconductivity} would get contribution only from $\omega =0$ and also from the Fermi surface when $\tau$ is large, we can approximate
\begin{align}\label{zerotempfermidirac}
\rho(\omega,\mathbf{k})^2 \approx -2\pi b \delta(\epsilon_k) \times 2 \tau
\end{align}
where we have utilized \eqref{Eq:FLrho1} for the first factor, and then substituted $\omega = 0$ and $\epsilon_k =0$ in the second factor. It is then easy to see that Eq.~\eqref{dcconductivity} reduces to
\begin{align}
    \sigma_{\rm dc} = \frac{e^2}{m^2}\tau \frac{b}{4\pi}\times 2 \times m\int_0^\infty {\rm d}y\,\, y \delta (y - 2m E_F)
\end{align}
with $y = k^2$ (note we do the $\omega$ integral first). Therefore
\begin{align}
    \sigma_{\rm dc} = e^2\frac{E_F}{\pi} b\tau = \frac{n e^2}{m}b \tau,
\end{align}
since $n/m = E_F/\pi$ for the Fermi liquid. 

We can similarly proceed to evaluate  Eq.~\eqref{hallconductivity} and obtain the Hall conductivity.

We use the approximation \eqref{Eq:FLrho1} for $\rho(\omega,\mathbf{k})$. Since the integral then gets contribution from $\omega =0$ and the Fermi surface only, we can use 
\begin{align}
\frac{\partial}{\partial k_x} {\rm Re} G_R (\omega,\mathbf{k}) \approx -\,b^{-1} \tau^2 \frac{\partial \,\epsilon_k}{\partial k_x} = - \,b^{-1} \tau^2 \frac{k_x}{m}
\end{align}
Therefore,
\begin{align*}
\sigma_{\rm H} = \frac{\omega_{c} e^2 \tau^2}{2\pi^2} \int {\rm d}^2 k \,\,\delta(\epsilon_k) \left(\frac{k_x}{m}\right)^2
\end{align*}
Now
\begin{align*}
& \int {\rm d}^2 k \,\,\delta(\epsilon_k) \left(\frac{k_x}{m}\right)^2  =\frac{2}{m}  \int {\rm d}^2 k \,\,\delta(k_x^2 + k_y^2 - 2mE_F ) \,{k_x}^{2}\\
& \hspace{3cm}=\frac{2}{m}  \int_{0}^{\infty}{\rm d}k \,\,k^3\,\delta(k^2 - 2mE_F )\\
&\hspace{5cm}\times \int_{0}^{2\pi}{\rm d}\theta \,{\rm cos}^{2}\theta\\
& \hspace{3cm}=\frac{\pi}{m}  \int_{0}^{\infty}{\rm d}y \,\,y\,\delta(y - 2mE_F )\\
& \hspace{3cm}= 2\, \pi E_{F}.
\end{align*}
Therefore,
\begin{align}
\sigma_{\rm H} =  \frac{E_{F}}{\pi} e^2 \omega_{c} \tau^2 = \frac{ne^2}{m} \omega_{c} \tau^2.
\end{align}
The Hall coefficient of the Fermi liquid is
\begin{equation}
    R = \frac{\sigma_{\rm H}}{B\sigma_{\rm dc}^2} = \frac{1}{n e}\times\frac{1}{b^2}.
\end{equation}

\bibliography{cond-refs}

\end{document}